# Dominance as an Indicator of Rapport and Learning in Human-Agent Communication

**Amanda Buddemeyer**
amb467@pitt.edu

**Xiaoyi Tian**
xiaoyi-tian@pitt.edu

**Erin Walker**
eawalker@pitt.edu

School of Computing and Information
University of Pittsburgh, Pittsburgh PA 15260

## Abstract

Power dynamics in human-human communication can impact rapport-building and learning gains, but little is known about how power impacts human-agent communication. In this paper, we examine dominance behavior in utterances between middle-school students and a teachable robot as they work through math problems, as coded by Rogers and Farace's Relational Communication Control Coding Scheme (RCCCS). We hypothesize that relatively dominant students will show increased learning gains, as will students with greater dominance agreement with the robot. We also hypothesize that gender could be an indicator of differences in dominance behavior. We present a preliminary analysis of dominance characteristics in some of the transactions between robot and student. Ultimately, we hope to determine if manipulating the dominance behavior of a learning robot could support learning.

## 1 Introduction

As virtual agents are employed in increasingly sophisticated capacities, it becomes imperative that we understand how the complex fabric of human-human communication translates to human-agent communication. A large body of evidence indicates that human users automatically and implicitly assign social characteristics to agents (Fong et al., 2003; Leyzberg et al., 2011), which may extend to concepts of power. Lucas et al. (2019) found that agents have greater informational influence but less normative influence on human subjects than other humans, indicating that human-agent power relationships work differently than human-human power relationships. Li et al. (2015) found that an observer's perception of a robot's dominance or submission can impact the observer's valence toward the robot, indicating that the power relationship might impact interactions between user and agent. Understanding how power can impact human-agent conversation might be key to building successful human-agent relationships.

While power is fundamentally important in understanding relationships and communication between humans (Dunbar and Burgoon, 2005; Burgoon and Hale, 1984; Russell, 1938), humans may relate differently to agents than to other humans. Between human conversational partners, a speaker's ability to manage social status and appropriately support face claims has been linked to valence and rapport-building (Spencer-Oatey, 2007), and rapport-building has been linked to positive educational and goal outcomes (Zhao et al., 2014). We build on this knowledge about human-human communication, in combination with previous work on similarities and differences in human-agent communication, to study how verbal assertions of or requests for dominance can be predictive of learning.

As we'll discuss in further detail in the "Background" section, gender can play an important role in power dynamics between human communicators. Gender-based differences in communication styles and goals may play a role in human-agent dominance behavior as well as any resulting educational effects.

In this work we propose two measures to characterize dyadic conversations between students and a robotic learning companion as they pursue an educational goal. The *control* score indicates how likely the student was to assert dominance over the robot rather than request dominance (i.e., behave submissively) or act neutrally. The *agreement* score indicates whether the two speakers were in agreement about dominance roles rather than competing for roles or behaving neutrally. With respect to these categories, we propose three research questions:

1. **Is the agreement score predictive of rapport-building and/or learning outcomes?** Using Brown and Levinson (1987)'s definition of face as the "public self-image that every member wants to claim for himself", dominance agreement may be a type of mutual support for a face claim. Given previous work linking face support with rapport-building, we hypothesize that agreement will correlate to rapport and learning gain.

2. **Is the control score predictive of rapport-building and/or learning outcomes?** We hypothesize that dominance will correlate to rapport and learning gains based on previous work correlating human authority with learning outcomes and correlating dominance with positive valence toward robots. It's noteworthy that the robot examined in this work is positioned as a teachable agent who learns with and from the student, meaning it has a relatively low social position. The control score in this case may be correlated to the agreement score in that students who are more dominant may be naturally complementary with the robot.

3. **Is the gender of the user predictive of the control or agreement score?** We hypothesize that female users will have greater agreement while male users will assert greater control.

This paper will include a preliminary analysis of results for a subset of the students involved in the experiments. The next step will be to compile results to answer the questions outlined above. Ultimately, it may be possible to manipulate an agent's behavior to optimize the dominance behavior of the user. Burgoon and Dunbar (2000) discuss social dominance as an inherently dyadic phenomenon that can be manipulated in an individual by changing her co-speaker and/or situation, which indicates that an agent could be designed with dominance strategies that favor control and/or agreement.

## 2 Background

Power and dominance have been defined differently in various works, but for the purpose of this paper we will use Burgoon et al. (1998)'s definitions. Power, broadly, is the ability to impact the behavior of others via social means. Dominance is a dyadic assertion of power. When an assertion is successful then it pairs with submission, a dyadic *request* for dominance. While power is unlikely to change over the course of a conversation, a single speaker is likely to both assert and request dominance at various times depending on the context.

Previous work has examined the specific impacts of power on human relationships with and perceptions of robots or agents. Li et al. (2015) found that human observers perceived a robot to be less trustworthy and socially attractive when it displayed dominant behavior. The study participants were observing interactions between a robot and an actor or actress rather than interacting with the robot themselves, which could be quite different from the scenario in the study described here. However, the authors' results may suggest a relationship between dominance behavior and a user's valence toward a robot. Hashemian et al. (2018) and Hashemian (2019) propose that modeling realistic social power relationships will make for more believable and more persuasive agents.

Some aspects of a human-agent power relationship could mirror outcomes observed in human-human power relationships. Howley et al. (2011) found that, in a group study context, a student's authoritativeness was predictive of his or her learning gains. This study examined authority as a monadic power behavior, while dominance is a dyadic power behavior. Still, the authors' coding scheme for authoritativeness had a fair amount of overlap with our dominance coding scheme, and the authors discuss the impact of using authoritative utterances to position themselves within the group in a manner that is similar to assertions of or requests for dominance. If the findings of Howley et. al. extend to human-agent relationships, it's likely that a student's dominance over a robotic learning companion is predictive of learning.

Gender can also shape power relationships because female and male speakers tend to communicate with different goals and styles. Lakoff (1975)'s classic work on gender, power, and language presents female communication styles as "low-power," or designed to conform to social pressures for women to be less assertive or direct. In contrast, Bradac and Mulac (1995) found that, while women tend to communicate less directly than men, these differences are not necessarily tied to power. In a survey of publications related to gender and discourse, Bucholtz (2003) points out that findings of female indirectness can be Western-centric, but her findings do support men as more dominant communicators, while women are more likely to use their communication to maintain relationships. Similarly, Tannen et al. (1990) found that women are more likely to prioritize rapport-building in conversation while men are more likely to speak to build or maintain social status. This suggests that male speakers may tend to have a higher control score in conversation while women may tend to have a higher agreement score.

## 3 Experiments

We examined transcripts from a study using a NAO robot (https://www.softbankrobotics.com/us/NAO) as a teachable agent. As explained by Lubold et al. (2019), "Teachable agents are pedagogical agents that employ the 'learning-by-teaching' strategy, which facilitates learning by encouraging students to construct explanations, reflect on misconceptions, and elaborate on what they know." Students were asked to spend 30 minutes teaching the robot to solve problems involving proportions, equations, and ratios using spoken language and a touch-screen interface on a Microsoft Surface. The dialogue system used Google Speech API for automatic speech recognition (ASR) in real time, but the results from this paper are based on a human transcription of the audio recordings from Rev (https://www.rev.com/).

The study involved 48 students between the ages of 12 and 14, 26 female and 22 male. The NAO robot in this study was named "Emma." This study involved three conditions: (1) a non-social condition, (2) a non-social condition with acoustic entrainment, and (3) a social condition. Each participant was given a math pre-test and post-test to show learning gains. They also completed a post-survey with questions used to calculate a rapport score. More information on this study can be found in (Lubold et al., 2019).

## 4 Coding Scheme

For our examination of dominance behavior in the study transcripts, we used Rogers and Farace (1975)'s Relational Communication Control Coding Scheme (RCCCS), which involves three levels of coding. First, each conversational turn is given a numeric code based on message format (assertions, questions, talk-over, etc.) and response mode (sup-port, non-support, instruction, topic change, etc.). It's noteworthy that this coding is inherently dyadic - a particular utterance might be coded differently depending on the other speaker's previous utterance. For example, the utterance "the answer is three" might have a numeric code of 14 (assertion, answer) if the speaker used this message to answer a question, 13 (assertion, extension) if it were simply an observation, or 43 (talk-over, extension) if the speaker interrupted the other speaker.

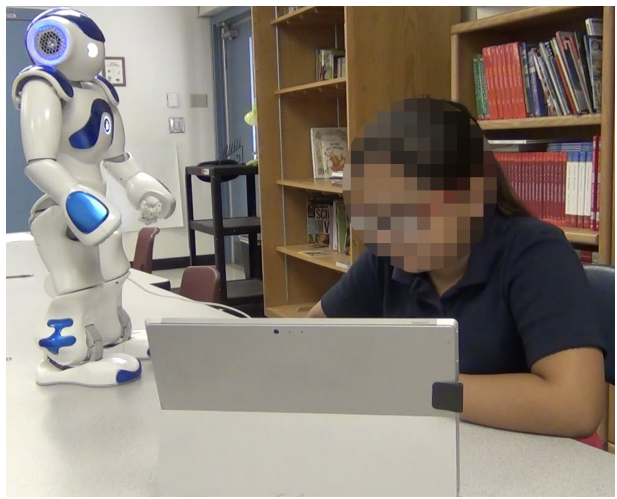

Figure 1: The NAO robot with a student.

For the second coding level, the numeric codes are translated to a control code: "one-up" ($\uparrow$), "one-down" ($\downarrow$), or "one-across" ($\rightarrow$), depending on whether the speaker is attempting to exert relational control, submitting to relational control by the co-speaker, or neither. Table 1 shows how some of the more common numeric codes translate to control codes. Finally, each pair of turns is then grouped into an agreement code: complementary transactions ($\uparrow\downarrow$ or $\downarrow\uparrow$) where both speakers agree about their dominant/submissive pairing, symmetrical transactions ($\uparrow\uparrow$, $\downarrow\downarrow$, or $\rightarrow\rightarrow$) where the speakers do not agree, and transitory transactions where exactly one of the turns is one-across and the speakers can't be said to either agree or disagree.

We made one significant change to the RCCCS coding to account for pedagogical questions. Pedagogical questions, as defined by Yu et al. (2019) are "questions asked by the person who knows the answer (or might reasonably expected to know the answer) to someone who may not know the answer, with the goal of eliciting learning." Lexically and syntactically, a pedagogical question might be indistinguishable from an information-seeking question; the underlying intent may be inferred by the roles of the interlocutors. For example, a student asking a teacher, "what is five plus eight" may be seeking information while a teacher asking a student the same question is most likely asking pedagogically. The RCCCS was not developed to specifically cover pedagogical dialogue and does not explicitly handle this type of question. Questions that are asked to provide assistance are coded as "one-down." Questions issued as instructions, orders, or as an answer to another question are coded as "one-up." To avoid confusion or ambiguity, we created a special code for pedagogical questions, which translates to a "one-up" code. This code was only used for the student because the student was assigned to the "tutor" role in the conversation. All of the robot's questions were presumed to be information-seeking.

| Message Format \ Response Mode | 1 (Support) | 2 (Non-support) | 3 (Extension) | 4 (Response) |
| --- | --- | --- | --- | --- |
| 1 (Assertion) | ↓ | ↑ | → | ↑ |
| 2 (Question) | ↓ | ↑ | ↓ | ↑ |
| 4 (Talk-over) | ↓ | ↑ | ↑ | ↑ |

Table 1: Partial table translating some of the more common RCCCS numerical codes to control codes. The full RCCCS has five message format codes and ten response mode codes. We added an additional response mode for pedagogical questions that is always coded "one-up" (↑).

Emma was designed to be a teachable robot, which implies a lower social position than a tutor. Much of what it is programmed to say could be categorized as questions and/or statements of agreement or support, all of which tend to be coded as "one-down" in the RCCCS. Thus, we expect to see some correlation between dominance behavior and agreement behavior.

## 5 Discussion and Results

Two independent coders coded transcripts for 15 students from the "Emma" study with a Cohen's Kappa of .819 on the numerical coding of 45.0% of the conversational turns. One student was excluded due to extensive technical problems with the robot. For the remaining 14 students, we coded a total of 1578 conversational turns, averaging 113 turns per conversation (52 average turns for each user, 61 average for the robot for each conversation) with a standard deviation of 16.6 turns per conversation. For each participant, we calculated a **control score** that is the percentage of the participant's total conversational turns that were coded as "one-up," or dominant. This score excludes the robot's conversational turns. For each participant, we also calculated an **agreement score** that is the percentage of transactions (pairs of conversational turns between the robot and the participant) that were coded as complementary (↑↓ or ↓↑). The scores are displayed in Table 2.

| Part. | Gender | Control Sc. | Agreement Sc. |
| --- | --- | --- | --- |
| 1 | male | 0.7045 | 0.6705 |
| 2 | female | 0.7647 | 0.6863 |
| 3 | female | 0.7500 | 0.6818 |
| 4 | male | 0.6923 | 0.6410 |
| 5 | female | 0.7174 | 0.5761 |
| 6 | male | 0.6491 | 0.5446 |
| 7 | female | 0.5536 | 0.4821 |
| 8 | female | 0.7213 | 0.5984 |
| 9 | female | 0.8444 | 0.7444 |
| 10 | female | 0.5744 | 0.5532 |
| 11 | male | 0.6471 | 0.5392 |
| 12 | female | 0.5522 | 0.4925 |
| 13 | male | 0.4630 | 0.4074 |
| 15 | male | 0.5738 | 0.4426 |

Table 2: Control and agreement scores for a subset of participants. The control score is the percentage of the user's conversational turns that are "one-up." The agreement score is the percentage of transactions (pairs of turns between the user and robot) that were complementary.

The 14 students have a mean control score of 0.6577 and a median control score of 0.6707. This is supportive of our hypothesis that students would tend to assert control when interacting with Emma given Emma's low power positioning as a teachable agent. By contrast, Emma had only 61 "one-up" turns over a total of 855 turns conversing with the 14 students, giving the robot an aggregate control score of 0.0713. Table 2 also provides some support for a correlation between a student's control score and agreement score. The mean agreement score is 0.5757 and the median is 0.5646. Given the fact that so few of Emma's utterances were "one-up," it makes sense that a participant's "one-up" messages will be more likely to pair with a

"one-down" message from Emma to create a complementary transaction (↑↓ or ↓↑).

Using the RCCCS, information-seeking questions tend to be coded as "one-down", as do statements of support or agreement like "yes" or "great." Instructions and direct informative answers to questions are always coded as "one-up." Conversations with Emma often follow a pattern of Emma asking a question or making a supportive statement (both of which are typically coded as "one-down") followed by the participant providing an answer or instruction (coded as "one-up"). This pattern is illustrated below in the conversation excerpt between Emma and User 8:

↓ **Emma:** This time I have the unit rate and I know how much battery I use in one hour. So I know three times one is three. What do I do next?

↑ **User 8:** So something has to equal three hours. So we need to figure out what would equal three hours, so we would use one because one times something equals three.

↓ **Emma:** So I have three over forty and I multiply it by three because I have three hours instead of one? And that's it?

↑ **User 8:** So you would do one times three.

Emma asked two questions, which User 8 answered directly with instructions about how to solve the problem. Both of User 8's messages were coded "one-up," which complements Emma's "one-down" messages. This pattern works similarly when the user asks pedagogical questions, such as this excerpt from User 15:

↓ **Emma:** This time I have the unit rate and I know how much battery I use in one hour. So I know three times one is three. What do I do next?

↑ **User 15:** One times three equals three. Three over 40 [inaudible 00:02:08] times three equals?

One reason that a student might receive a lower control or agreement score is by minimally participating. The RCCCS considers a response of "yes" without elaboration as "support" and this is coded as "one-down." Similarly, a response of "okay" is considered an "extension" (a neutral continuation of the conversation) and is coded as "one-across." Either type of response from a student will reduce the control score, and will also reduce the agreement score if paired with a message from Emma that is also "one-down." This excerpt from User 13's transcript provides an example:

↓ **Emma:** I'm thinking. So if I look at the ratio of one to three. Three is three times more. Right?

↓ **User 13:** Yes.

↓ **Emma:** I was thinking that three times more than three would be nine. What do you think?

↓ **User 13:** Yes.

In this case, the student's low-control behavior may be indicative of a lack of engagement. Because all of the messages in the above excerpt are coded as "one-down," all of the transactions (pairs of messages) disagree with each other (↓↓) and are considered symmetrical. This means that they lower the agreement score for this conversation. A transaction in which both messages are coded as one-up (↑↑) would also be symmetrical.

Among the fourteen students examined in this work, the overwhelming majority of "one-down" control codes were due to brief agreement responses such as "yes" or "correct." We do see some non-pedagogical questions from students that are coded as "one-down," the most common of which is simply asking the robot to repeat itself. We also see some words of encouragement or more elaborated statements of support, such as this exchange with User 8:

↓ **Emma:** I was thinking that three times more than three would be nine. What do you think?

↓ **User 8:** Three times three equals nine.

→ **Emma:** Okay I will put the answer is nine.

↓ **User 8:** Good job.

In other cases, a symmetrical transaction may be less due to the student's behavior than the robot's. Earlier in the same exchange between the robot and User 13, Emma says "okay," which is coded as "one-across." This utterance is neither dominant nor submissive:

↓ **Emma:** Can you give me a hint to figure out how much paint I need for my body?

↓ **User 13:** Yes.

→ **Emma:** Okay.

↑ **User 13:** You have to add six and three.

The "okay" message is part of two transactions: the pair of messages that begins when User 13 says "Yes" (↓→) and the pair of messages that begins with Emma's "Okay" message and ends with the user instructing Emma to add six and three (→↑). Both transactions are coded as transitory because they neither agree nor disagree. A transitory transaction will also lower the agreement score.

The control and agreement scores are also impacted when the robot utters a message coded as "one-up." Of Emma's 61 "one-up" turns, exactly 14 (one per student) are due to her initiating the conversation. The RCCCS almost always codes an initiation statement as "one-up." While the robot described in this work always initiates the conversation verbally, this initiation does not occur until the student has begun using the tablet interface. It is not clear whether human users react to a verbal initiation from the robot in this case as they would to a similar initiation from another human. One of Emma's "one-up" turns is due to her changing the topic. Another 22 "one-up" turns are due to Emma issuing an instruction or order. Some of these are problem-related, such as "Help me figure out an equation for Zach!" Others relate to moving through the problem set, such as "Press the start teaching button and we can get started!" The remaining 24 "one-up" turns were caused by two phenomena related to technical problems with the robot: **disconfirmation** (15 turns), which is always coded as "one-up" in the RCCCS, and **talk-overs** (nine turns), which are often but not always coded as "one-up."

A disconfirmation occurs if one speaker utters something that requests a response and the other speaker ignores that request. Emma utters a disconfirmation if user asks Emma a question not handled by the dialog system:

↑ **User 15:** One battery equals how many hours?

↑ **Emma:** Okay that makes sense.

A talk-over is an interruption, which might occur if the participant speaks softly or Emma otherwise cannot detect that the person is speaking. It is unclear whether human speakers display the same dominance behavior in reaction to a robot/agent disconfirmation or talk-over as they would to a human co-speaker making the same type of utterance.

These 14 participants include six male and eight female students, but there are no apparent patterns between gender and either control score or agreement score. Viewing the results from the full set of participants for both studies may provide additional information.

The next steps for this study involve coding the remaining participants in the NAO robot studies and investigating possible correlations between control and agreement score versus learning gain, rapport score, and gender. If the hypotheses of this paper are supported, a logical next step would be to design an intervention to determine a causal relationship and to see how students react to a robot or agent that can adapt its dominance behavior.

Previous work could provide a direction for such an intervention. Worgan and Moore (2011) propose a strategy of observing entrainment in prosodic and spectral features of speech to determine the relative social power of the interlocutors. Danescu-Niculescu-Mizil et al. (2012), like Worgan et. al., proposes using entrainment to detect power differences, but the authors focus on lexical entrainment rather than entraining on prosodic features. Danescu-Niculescu-Mizil et al. (2013) formulated a computational approach to detecting politeness in online conversations and found a negative correlation between politeness and the relative social status of the speaker. While neither social power nor social status are the same as dominance, they are sufficiently related to suggest a promising research direction.

Power and dominance behaviors are a major contributor to human relationships and human-human communication; research indicates that they might be important in human-agent relationships and communication as well. The results of this study may suggest an important research direction to improve rapport-building with and learning gains from pedagogical agents.

### Acknowledgments

This work was supported by National Science Foundation Award Nos. DRL-1811610 and IIS-1637809. We would also like to thank Dr. Nichola Lubold and Dr. Rob Voigt for their invaluable help.

# References

James J. Bradac and Anthony Mulac. 1995. Women's style in problem solving interaction: Powerless, or simply feminine? In Pamela J. Kalbfleisch and Michael J. Cody, editors, *Gender, Power, and Communication in Human Relationships*, chapter 4, pages 83–104. Lawrence Erlbaum Associates, Inc., Hillsdale, New Jersey.

Penelope Brown and Stephen C Levinson. 1987. *Universals in language usage: Politeness phenomena*. Cambridge University Press, Cambridge.

Mary Bucholtz. 2003. Theories of discourse as theories of gender: Discourse analysis in language and gender studies. In Janet Holmes and Miriam Meyerhoff, editors, *The Handbook of Language and Gender*, chapter 2, pages 43–68. Blackwell, Oxford.

Judee K Burgoon and Norah E Dunbar. 2000. An interactionist perspective on dominance-submission: Interpersonal dominance as a dynamic, situationally contingent social skill. *Communications Monographs*, 67(1):96–121.

Judee K Burgoon and Jerold L Hale. 1984. The fundamental topoi of relational communication. *Communication Monographs*, 51(3):193–214.

Judee K Burgoon, Michelle L Johnson, and Pamela T Koch. 1998. The nature and measurement of interpersonal dominance. *Communications Monographs*, 65(4):308–335.

Cristian Danescu-Niculescu-Mizil, Lillian Lee, Bo Pang, and Jon Kleinberg. 2012. Echoes of power: Language effects and power differences in social interaction. In *Proceedings of the 21st international conference on World Wide Web*, pages 699–708.

Cristian Danescu-Niculescu-Mizil, Moritz Sudhof, Dan Jurafsky, Jure Leskovec, and Christopher Potts. 2013. A computational approach to politeness with application to social factors. In *Proceedings of ACL*. ACL.

Norah E Dunbar and Judee K Burgoon. 2005. Perceptions of power and interactional dominance in interpersonal relationships. *Journal of Social and Personal Relationships*, 22(2):207–233.

Terrence Fong, Illah Nourbakhsh, and Kerstin Dautenhahn. 2003. A survey of socially interactive robots. *Robotics and autonomous systems*, 42(3-4):143–166.

Mojgan Hashemian. 2019. Persuasive social robots using social power dynamics. In *Proceedings of the 18th International Conference on Autonomous Agents and MultiAgent Systems*, pages 2408–2410. International Foundation for Autonomous Agents and Multiagent Systems.

Mojgan Hashemian, Rui Prada, Pedro A Santos, and Samuel Mascarenhas. 2018. Enhancing social believability of virtual agents using social power dynamics. In *Proceedings of the 18th International Conference on Intelligent Virtual Agents*, pages 147–152. ACM.

Iris Howley, Elijah Mayfield, and Carolyn P Rosé. 2011. Missing something? authority in collaborative learning. In *Proceedings of the 9th International Conference on Computer Supported Collaborative Learning*, pages 336–373.

Robin Tolmach Lakoff. 1975. *Language and Woman's Place: Text and Commentaries (Studies in Language and Gender)*. Harper & Row, New York.

Dan Leyzberg, Eleanor Avrunin, Jenny Liu, and Brian Scassellati. 2011. Robots that express emotion elicit better human teaching. In *Proceedings of the 6th international conference on Human-robot interaction*, pages 347–354. ACM.

Jamy Li, Wendy Ju, and Cliff Nass. 2015. Observer perception of dominance and mirroring behavior in human-robot relationships. In *2015 10th ACM/IEEE International Conference on Human-Robot Interaction (HRI)*, pages 133–140. IEEE.

Nichola Lubold, Erin Walker, Heather Pon-Barry, and Amy Ogan. 2019. Comfort with robots influences rapport with a social, entraining teachable robot. In *International Conference on Artificial Intelligence in Education*.

Gale M Lucas, Janina Lehr, Nicole Krämer, and Jonathan Gratch. 2019. The effectiveness of social influence tactics when used by a virtual agent. In *Proceedings of the 19th ACM International Conference on Intelligent Virtual Agents*, pages 22–29. ACM.

L Edna Rogers and Richard V Farace. 1975. Analysis of relational communication in dyads: New measurement procedures. *Human Communication Research*, 1(3):222–239.

Bertrand Russell. 1938. *Power: A new social analysis*. W. W. Norton, New York.

Helen Spencer-Oatey. 2007. Theories of identity and the analysis of face. *Journal of pragmatics*, 39(4):639–656.

Deborah Tannen et al. 1990. *You just don't understand: Women and men in conversation*. Morrow New York.

Simon F Worgan and Roger K Moore. 2011. Towards the detection of social dominance in dialogue. *Speech Communication*, 53(9-10):1104–1114.

Yue Yu, Elizabeth Bonawitz, and Patrick Shafto. 2019. Pedagogical questions in parent–child conversations. *Child development*, 90(1):147–161.

Ran Zhao, Alexandros Papangelis, and Justine Cassell. 2014. Towards a dyadic computational model of rapport management for human-virtual agent interaction. In Timothy Bickmore, Stacy Marsella, and Candy Sidner, editors, *Intelligent Virtual Agents*, volume 8109 of *Lecture Notes in Computer Science*. Springer-Verlag, Vienna, Austria.